\documentclass[%
 aip,
 jcp,
 amsmath,amssymb,
reprint,%
]{revtex4-1}

\usepackage{graphicx}
\usepackage{dcolumn}
\usepackage{bm}
\usepackage{xcolor}
\usepackage{braket}
\usepackage{algorithm}
\usepackage{algpseudocode}
\usepackage{orcidlink}
\DeclareMathOperator{\Tr}{Tr}

\makeatletter
\newcommand{\thickhline}{%
    \noalign {\ifnum 0=`}\fi \hrule height 1pt
    \futurelet \reserved@a \@xhline
}
\newcolumntype{"}{@{\hskip\tabcolsep\vrule width 1pt\hskip\tabcolsep}}
\makeatother

\makeatletter
\def\@email#1#2{%
 \endgroup
 \patchcmd{\titleblock@produce}
  {\frontmatter@RRAPformat}
  {\frontmatter@RRAPformat{\produce@RRAP{*#1\href{mailto:#2}{#2}}}\frontmatter@RRAPformat}
  {}{}
}%
\makeatother

\begin{document}

\preprint{AIP/123-QED}

\title[Positivity Preserving Density Matrix Minimization at Finite Temperatures via Square Root]{Positivity Preserving Density Matrix Minimization at Finite Temperatures via Square Root}
\author{Jacob Leamer}
\affiliation{Department of Physics and Engineering Physics, Tulane University, 6823 St. Charles Ave., New Orleans, LA 70118, USA}
\author{William Dawson}
\affiliation{ 
RIKEN Center for Computational Science, Kobe, Hyogo,
650-0047, Japan.
}%
\author{Denys I. Bondar}
\affiliation{Department of Physics and Engineering Physics, Tulane University, 6823 St. Charles Ave., New Orleans, LA 70118, USA}

\date{\today}

\begin{abstract}
We present a Wave Operator Minimization (WOM) method for calculating the Fermi-Dirac density matrix for electronic structure problems at finite temperature while preserving physicality by construction using the wave operator, i.e., the square root of the density matrix. WOM models cooling a state initially at infinite temperature down to the desired finite temperature. We consider both the grand canonical (constant chemical potential) and  canonical (constant number of electrons) ensembles.  Additionally, we show that the number of steps required for convergence is independent of the number of atoms in the system.  We hope that the discussion and results presented in this article reinvigorates interest in density matrix minimization methods.
\end{abstract}

\maketitle

\section{Introduction}\label{sec:intro}
The ability to determine the electronic structure is of critical importance for obtaining a proper understanding of the quantum behavior of  materials. One way of many~\cite{akimov_large-scale_2015} to achieve this is by computing the finite temperature, effective single-electron density matrix $\hat\rho$ (or Gibbs state) that statistically describes the system state in question
\begin{equation}
    \hat\rho(\beta) = \frac{\hat{I}}{\hat{I} + \exp[\beta(\hat{H}-\mu)]},
\end{equation}
where $\hat{H}$ is the $N$-dimensional Hamiltonian operator, $\hat{I}$ the identity operator, and $\beta=1/(kT)$ is the inverse temperature of the system. Once $\hat\rho$ has been determined, it can be used to obtain other information about the system such as the number of electrons $N_e = \Tr(\hat\rho)$ or the electronic energy $E = \Tr(\hat{H}\hat\rho)$, which in general depends upon $\hat\rho$.  These quantities can then be used to obtain other observables of interest like the specific heat $C_v = \frac{d E}{d T}$ where $T$ is the temperature of the system.

Unfortunately, conventional methods for obtaining $\hat\rho$ are known to scale poorly with the size of the system in question.  This is because they generally rely on diagonalization of $\hat{H}$ or the multiplication of dense matrices, both of which scale as $O(N^3)$.  As such, a lot of effort has been directed toward the development of new methods for solving these problems that scale linearly with $N$~\cite{goedecker_1999,Bowler_2012}. 
These methods have proven useful in performing Density Functional Theory calculations~\cite{ratcliff-2016-challenges,Dawson2022}, including \textit{ab initio} molecular dynamics~\cite{cawkwell_energy_2012,Schade2022,Niklasson2017,Negre2023}; this success has prompted the general development of a number of open-source libraries~\cite{Borstnik2014, lin_fast_2011, lin_selinv---algorithm_2011, jacquelin_pselinvdistributed_2017, Rubensson2022, Bock2018, mohr_efficient_2017, dawson_massively_2018}.

Despite their usefulness, the vast majority of the linear-scaling methods are focused on obtaining the density matrix describing zero-temperature states, i.e., the ground state of the system.  However, in recent years there has been a great deal of interest in the development of methods for calculating $\hat\rho$ at finite temperature~\cite{PhysRevB.51.9455, niklasson_2003, Niklasson2008, pratapa_2016, aarons_perspective_2016, mniszewski_linear_2019, mohr_efficient_2017, aarons_electronic_2018}, including stochastic methods~\cite{PhysRevB.97.115207, Baer2022}. 
Density Functional Theory calculations at finite temperature~\cite{PhysRev.137.A1441,10.1103/PhysRevA.37.2809,PhysRevA.37.2821} are a requirement for the study of warm dense matter~\cite{10.1007/978-3-319-04912-0_2}, may be used to describe static correlation~\cite{Chai2012, Grimme2015, Filatov2015}, can accelerate convergence of self-consistent field methods~\cite{Rabuck1999}, or be employed for studying chemical reactions~\cite{mniszewski_linear_2019}.
To contribute to this growing body of work, we present methods for calculating $\hat\rho$ at a finite temperature for electronic systems that are in the grand canonical or canonical ensembles.

Our methods are based on the \emph{Wave Operator Minimization} (WOM), where the wave operator refers to the square root of the density matrix.  The core idea of this method is that the square root of the Fermi-Dirac state is the solution to an initial value problem. This is conceptually similar to the approach utilized in density matrix minimization methods~(DMM)~\cite{daw_model_1993, challacombe_simplified_1999, li_density-matrix_1993, lai_localized_2016, lai_density_nodate, Arita2014, Daniels1997, Millam1997, Bowler1999, Challacombe1999, Shao2003, Helgaker2000,Larsen2001,Salek2007,nunes_generalization_1994, hernandez_linear-scaling_1996}, but casting the initial value problems in terms of the square root of $\hat\rho$ yields solutions that are manifestly positive and Hermitian -- two crucial physicality constraints.

The formalism of the square root of a density matrix -- also known as the wave operator -- has been used for foundational research~\cite{uhlmann_parallel_1986, wlodarz_quantum_1994, reznik_unitary_1996, gheorghiu-svirschevski_nonlinear_2001, beretta_nonlinear_2005, beretta_nonlinear_2006, yahalom_square-root_2006, tronci_momentum_2019, mccaul2023wave}. The wave operator also has turned out to be as an efficient computational tool for both Markovian and non-Markovian open quantum system dynamics~\cite{joubert-doriol_non-stochastic_2014}.  To best of our knowledge, the current paper would perhaps be the first deployment of the density matrix square root for electron structure calculations. Note that the square root of the electronic density, which is contained in the diagonal elements of a density matrix, has been used as a numerical tool for orbital free density functional theory~\cite{levy_exact_1984, levy_exact_1988, flores_differential_1992}.

The rest of the paper is organized as follows. In Sec.~\ref{sec:fin_temp}, we introduce the WOM method using the Gibbs state and contrast it with a DMM-like approach.  Sections~\ref{sec:gc_fin_temp} and \ref{sec:c_fin_temp} discuss the application of WOM to fermionic systems that obey the statistics of the grand canonical and canonical ensembles, respectively.  We discuss the algorithmic implementation of our methods and demonstrate its effectiveness by considering a models of bulk aluminum and silicon of different system sizes in Sec.~\ref{sec:results}.

\section{From finite Temperature DMM to WOM}\label{sec:fin_temp}
As mentioned previously, the difficulty with calculating $\hat\rho$ is that the computational costs typically scale as $O(N^3)$, which makes it infeasible for many systems of interest.  Approaches that are based on minimization, however, are able to achieve $O(N)$ scaling by using sparse matrix algebra computational kernels.  Fortunately, sparsity naturally arises in electronic structure calculations from Kohn's nearsightedness principle~\cite{kohn_1996, prodan_nearsightedness_2005}, which holds even for metals at finite temperatures.  Because of this, minimization-based approaches are a natural consideration for the development of new methods for solving electronic structure problems.

We were further motivated to consider an iterative approach for calculating $\hat\rho$ at finite temperatures by the method developed by Bloch~\cite{bloch_1932} to obtain the Gibbs state.
For such states 
\begin{equation}
\frac{d \hat\rho}{d \beta} = -\hat{H}\hat\rho.
\label{bloch_deriv}
\end{equation}
This equation along with the initial condition $\hat\rho(0) = \hat{I}$ (i.e. the identity matrix) can be used to construct an iterative method for finding $\hat\rho$ at any given temperature.  The problem with using Eq.~\eqref{bloch_deriv} is that numerical errors in the calculation may make $\hat\rho$ non-Hermitian. To avoid this issue, Bloch symmetrized Eq.~\eqref{bloch_deriv} in the following manner
\begin{equation}
    \frac{d \hat\rho}{d \beta} = -\frac{\hat{H}\hat\rho}{2} - \frac{\hat\rho\hat{H}}{2}.
    \label{bloch_sym_deriv}
\end{equation}

To elucidate the power of the Bloch method, let us prove that evolving $\hat\rho$ via Eq.~\eqref{bloch_sym_deriv} preserves not only Hermiticity but also the positivity of the density matrix. The representation of Eq.~\eqref{bloch_sym_deriv} via a finite difference approximation
\begin{align}
    \frac{\hat\rho(\beta+\delta\beta) - \hat\rho(\beta)}{\delta\beta} = -\frac{\hat{H}\hat\rho(\beta)}{2} - \frac{\hat\rho(\beta)\hat{H}}{2} + O(\delta\beta),
\end{align}
where $\delta\beta$ is a small inverse temperature increment, can be recast as
\begin{multline}
    \hat\rho_{n+1} = \left(\hat{I}-\frac{\delta\beta}{2}\hat{H}\right)\hat\rho_n\left(\hat{I}-\frac{\delta\beta}{2}\hat{H}\right)^\dag + O(\delta\beta^2).
    \label{pos_bloch}
\end{multline}
Here, $\hat\rho_n = \hat\rho(\beta)$ and $\hat\rho_{n+1} = \hat\rho(\beta+\delta\beta)$.  Because the r.h.s. of Eq.~\eqref{pos_bloch} is in the form $A\hat\rho_nA^\dag$, we know that this method also preserves the positivity of $\hat\rho$~\cite{horn_mat_analysis}.  As such, we can think of this iterative method as behaving like a quantum channel modeling cooling that maps $\hat\rho_n$ from a higher temperature state to a lower one, $\hat\rho_{n+1}$.  This approach is conceptually similar to the quenching approach utilized for the Fermi operator expansion method by Aarons and coworkers~\cite{aarons_electronic_2018}.  One key difference, however, is that the quenching Fermi operator approach does not explicitly preserve the positivity of $\hat\rho$, while the symmeterized derivative approach in Eq.~\eqref{bloch_sym_deriv} can be shown to preserve positivity.

Whereas the symmetrization approach generalizes to the Fermi-Dirac state, we found that it is easier to perform the calculations in terms of WOM. Let us introduce our method by evaluating the Gibbs state $\hat\rho(\beta)$. The key idea is to represent the density matrix as
\begin{align}
    \hat\rho(\beta) &= \hat\Omega(\beta)^{\dagger} \hat\Omega(\beta), \label{eq:OmegaGibbs}\\
    \hat\Omega(\beta) &= \exp(-\beta \hat{H}/2), \notag
\end{align}
where the square root of the Gibbs state $\hat\Omega(\beta)$ is the solution to the initial value problem
\begin{align}\label{eq:OmegaGibbsDiffEq}
    \frac{d}{d\beta} \hat\Omega(\beta) = -\frac{1}{2} \hat{H} \hat\Omega(\beta), 
    \qquad \hat\Omega(0) = \hat{I}.
\end{align}
The main advantage of the WOM method is that one can use \emph{any} numerical method for solving Eq.~\eqref{eq:OmegaGibbsDiffEq} and the resulting Gibbs state~\eqref{eq:OmegaGibbs} is manifestly positive. That is we are not restricted to specific finite difference methods, e.g., Eq.~\eqref{pos_bloch}.

\section{WOM for the grand canonical ensemble}\label{sec:gc_fin_temp}

In this section, we apply WOM to fermionic systems that are described by the Fermi-Dirac distribution in a non-orthonormal basis and whose statistical behavior is described by the grand canonical ensemble in which systems are allowed to exchange both energy and electrons with their environment.  For us, this means that we need to fix the value of the chemical potential $\mu$ before applying our method, which will allow the number of electrons $N_e$ to change.

Because we wish to consider a non-orthonormal basis, we need to introduce the overlap matrix $S$.  The overlap matrix is a square, positive-definite matrix that characterizes the overlap between the basis vectors used to describe a system.  For basis vectors $\ket{i}, \ket{j}$ the elements of the overlap matrix are given by the following expression
\begin{equation}
    S_{ij} = \left< i \middle| j \right>.
\end{equation}
We use $P$ and $H$ to denote the matrix representations of $\hat\rho$ and $\hat{H}$ respectively in the non-orthonormal basis. The elements of these matrices $P$ and $H$ are
\begin{align}
    P_{ij} &= \braket{i|\hat\rho|j}, &  H_{ij} &= \braket{i|\hat{H}|j}.
\end{align}
Using these representations, we can obtain the following expression for the Fermi-Dirac distribution in a general basis (see Appendix~\ref{app:NonOrthonormalFD} for details)
\begin{align}
    P(\beta) = S\left[I+\exp[\beta(S^{-1}H-\mu I)]\right]^{-1}.
    \label{non_orth_fd}
\end{align}
The corresponding WOM representation is 
\begin{align}
    P(\beta) &= \Omega(\beta)^\dag\Omega(\beta), \label{eq:sqrt_P}\\
    \Omega(\beta) &= S^{1/2}\left[I + \exp[\beta(S^{-1}H-\mu I)]\right]^{-1/2}, \notag
\end{align}
(see Appendix \ref{app:omega_proof} for proof) and the initial value problem to which $\Omega(\beta)$ is the solution is
\begin{align}
    \frac{d\Omega}{d\beta} &= -\frac{1}{2}\Omega\left[I-(S^{-1/2}\Omega)^2\right](S^{-1}H-\mu I), \label{eq:OmegaGC_deriv} \\
    \Omega(0) &= \sqrt{\frac{S}{2}}. \label{eq:OmegaGC_iv}    
\end{align}

Several DMM methods have been developed for handling systems in a non-orthogonal basis and with statistics described by the grand canonical ensemble~\cite{nunes_generalization_1994, hernandez_linear-scaling_1996}.  However, these methods are aimed at obtaining the ground state density matrix and thus rely heavily upon the McWeeney purification algorithm~\cite{mcweeny_recent_1960} to enforce idempotency and positivity on the resulting $P$.  Unlike the approaches used in many DMM methods~\cite{daw_model_1993, challacombe_simplified_1999, li_density-matrix_1993, nunes_generalization_1994, hernandez_linear-scaling_1996, palser_canonical_1998}, Eq.~\eqref{eq:sqrt_P} is shown to be explicitly positivity-preserving without the need for something like the McWeeney purification algorithm or an exponential parameterization~\cite{Shao2003,Helgaker2000,Larsen2001,Salek2007}.

\section{WOM for the Canonical Ensemble}\label{sec:c_fin_temp}

It is also possible to apply WOM to systems described by the canonical ensemble in which a system is only allowed to exchange energy with the environment.  This means that $N_e$ should remain constant as we iterate towards the desired temperature.  Note that in the non-orthogonal basis, $N_e = \Tr[S^{-1}P]$.  This constraint can be enforced by updating the value of $\mu$ with every step.  Thus, we now have $\mu = \mu(\beta)$ and Eq.~\eqref{non_orth_fd} becomes
\begin{align}
    P(\beta) &= S\left[I+\exp[\beta(S^{-1}H-\mu(\beta) I)]\right]^{-1} \label{eq:c_non_orth_fd}.
\end{align}
The corresponding form of $\Omega$ is
\begin{align}
    \Omega(\beta) &= S^{1/2}\left[I + \exp[\beta(S^{-1}H-\mu(\beta) I)]\right]^{-1/2}.
\end{align}
For brevity, we introduce
\begin{align}
    X &= \Omega\left[I-(S^{-1/2}\Omega)^2\right], \\
    A &= S^{-1}H.    
\end{align}
Now the derivative with respect to $\beta$ is
\begin{align}
    \frac{d\Omega}{d\beta} = -\frac{1}{2}X(A-\mu-\beta\frac{d\mu}{d\beta}). \label{eq:OmegaCDeriv}
\end{align}
To guarantee that $N_e$ is conserved while $\beta$ is increasing, we solve $\frac{dN_e}{d\beta} = \frac{d}{d\beta}\Tr{[S^{-1}\Omega(\beta)^\dag\Omega(\beta)]} = 0$ for the unknown $\mu + \beta\frac{d\mu}{d\beta}$ to obtain
\begin{align}
    \mu + \beta\frac{d\mu}{d\beta} = \frac{\Tr[AS^{-1}X^\dag\Omega + \Omega^\dag X S^{-1}A^{\dag}]}{\Tr[S^{-1}X^\dag\Omega + \Omega^\dag XS^{-1}]}. \label{eq:OmegaCSub}
\end{align}
Then we substitute Eq.~\eqref{eq:OmegaCSub} into Eq.~\eqref{eq:OmegaCDeriv} to construct the sought initial value problem
\begin{align}
    \frac{d\Omega}{d\beta} &= -\frac{1}{2}X\left(A-\frac{\Tr[AS^{-1}X^\dag\Omega + h.c.]}{\Tr[S^{-1}X^\dag\Omega + h.c.]}\right), \label{eq:OmegaSubbedCDeriv}\\
    \Omega(0) &= \sqrt{\frac{N_e}{N}}\sqrt{\frac{S}{2}} \label{eq:OmegaCIV},
\end{align}
where Eq.~\eqref{eq:OmegaCIV} has been constructed so that $\Tr[S^{-1}P(0)] = N_e$.  By solving this initial value problem, we can calculate $P(\beta) = \Omega(\beta)^\dag\Omega(\beta)$ while ensuring that $N_e$ is conserved.

A review of DMM methods reveals that the presence of $\frac{d\mu}{d\beta}$ in Eq.~\eqref{eq:OmegaCDeriv} appears to be an often overlooked subtlety. The methods that explicitly preserve $N_e$~\cite{daw_model_1993, palser_canonical_1998} make no mention of the importance of updating the value of $\mu$, while others~\cite{li_density-matrix_1993, challacombe_simplified_1999, nunes_generalization_1994, hernandez_linear-scaling_1996} focus on the grand canonical ensemble and only refer to the fact that conserving $\mu$ is easier than conserving $N_e$.  One method~\cite{Qiu_1994} that does mention the importance of updating $\mu$ does so by solving a third order polynomial as a part of a two-stage steepest descent algorithm.  In contrast, our approach of updating $\mu$ only requires evaluating the traces of matrices and taking their ratio.  It is also worth noting again that WOM preserves positivity explicitly without the need for the McWeeney purification algorithm used in the DMM methods.

The calculation of $\mu$ is another key difference between our minimization approach and the quenching Fermi operator expansion approach~\cite{aarons_electronic_2018}.  In the latter, $\mu$ is calculated using a root-finding approach that requires the inversion of matrices.  In contrast, our approach updates the value of $\mu$ to preserve $N_e$ while iterating $P$ to the desired temperature.

\section{Illustrations}\label{sec:results}

To illustrate WOM, we first consider a tight-binding model of aluminum.  The tight-binding model~\cite{slater_simplified_1954} has been used to great success in a number of solid state applications~\cite{maslov_nonorthogonal_2016, liu_three-band_2013} and thus serves as a useful benchmark for developing methods of solving electronic structure problems.  Aluminum is a simple metal that has been studied extensively~\cite{chuang_structure_2006, staszewska_many-body_2005}, which makes it convenient to obtain Hamiltonians for a variety of system sizes.  This in turn allows us to test the scaling of the adaptive step implementations of our methods while testing their effectiveness.  We used the DFTB+ library~\cite{hourahine_dftb_2020} with the matsci-0-3 parameters~\cite{frenzel_structural_2005, manzano_cement_2012} to generate the Hamiltonian and overlap matrices for our calculations.  To create the crystal structure for the DFTB+ calculations, we used the Atomic Simulator Environment~\cite{larsen_atomic_2017}. All of our calculations were done at the $\Gamma$ point only.

\begin{algorithm}[H]
    \caption{An adaptive implementation of WOM using the second-order Runge-Kutta method.}
    \label{adaptive_algorithm}
    \begin{algorithmic}
        \State Set $\delta\beta, \beta_f$ ($\beta_f$ the target inverse temperature)
        \State Set $\beta_i = 0$
        \State Set tolerance $\epsilon = 10^{-2}$
        \State Set early exit tolerance $\epsilon_2 = 10^{-4}$
        \State Set previous $\delta\beta$: $\delta\beta_p = 0$
        \State Set $H, S, \mu$
        \State Set $\Omega = \Omega_0$
        \While{$\beta_i < \beta_f$}
            \State Perform single step of RK1 to get $\Omega_c$
            \State Perform single step of RK2 to get $\Omega'$
            \State Calculate $e_n =  ||\Omega'-\Omega_c||_2$
            \While{$e_n > \epsilon$}
                \State $\delta\beta \gets \delta\beta  \sqrt{\frac{\epsilon}{e_n}}$
                \State Perform single step of RK2 to get new $\Omega'$
                \State Calculate $e_n = ||\Omega' - \Omega_c||_2$
            \EndWhile
            \State Calculate $d_n =  ||\Omega-\Omega'||_2$
            \State $\Omega \gets \Omega'$
            \State $\beta_i \gets \beta_i + \delta\beta$
            \State $\delta\beta \gets \delta\beta \sqrt{\frac{\epsilon}{e_n}}$
            \If{$d_n > \epsilon_2$}
                \State \textbf{break}
            \EndIf
        \EndWhile
        \State End
    \end{algorithmic}
\end{algorithm}

For the grand canonical method, we solve Eqs.~\eqref{eq:OmegaGC_deriv} and \eqref{eq:OmegaGC_iv}
using the average of the highest occupied and lowest unoccupied eigenvalues of $H$ as the value for $\mu$.  For the canonical method, we solve Eqs.~\eqref{eq:OmegaSubbedCDeriv} and \eqref{eq:OmegaCIV} while simultaneously solving Eq.~\eqref{eq:OmegaCSub} to update $\mu$.  For both cases, we end at a temperature of 3157 K. Both methods were implemented using an adaptive step~\cite{soderlind_adaptive_2006} second-order Runge-Kutta method with a tolerance $\epsilon = 10^{-2}$ (Algorithm~\ref{adaptive_algorithm}).  The code can be found online~\cite{leamer_git}.  The eigenvalue spectra of $P$ obtained by these methods for a 250 atom supercell of aluminum are shown in Fig.~\ref{fig:al_eigs} (absolute errors are plotted in Supplementary Information I).  The results of both methods coincide with the exact eigenvalue spectra obtained by solving Eq.~\eqref{non_orth_fd} or Eq.~\eqref{eq:c_non_orth_fd} directly.  We also compare electronic energies $E = 2\Tr[S^{-1}PS^{-1}H]$ obtained from our methods to those obtained via direct diagonalization for 16, 54, 128, and 250 atom supercells.  For the grand canonical method, we observe that the relative difference is between $0.0008\%$ and $0.0027\%$.  For the canonical method, the relative difference is between $0.0154\%$ and $0.0329\%$. For increased accuracy, the Runge-Kutta tolerance would need to be tightened and potentially combined with a higher order method. These results demonstrate the effectiveness of our methods in obtaining the Fermi-Dirac density matrices at finite temperatures for a tight-binding Hamiltonian in a non-orthogonal basis.

\begin{figure}
    \centering
    \includegraphics[width=\linewidth]{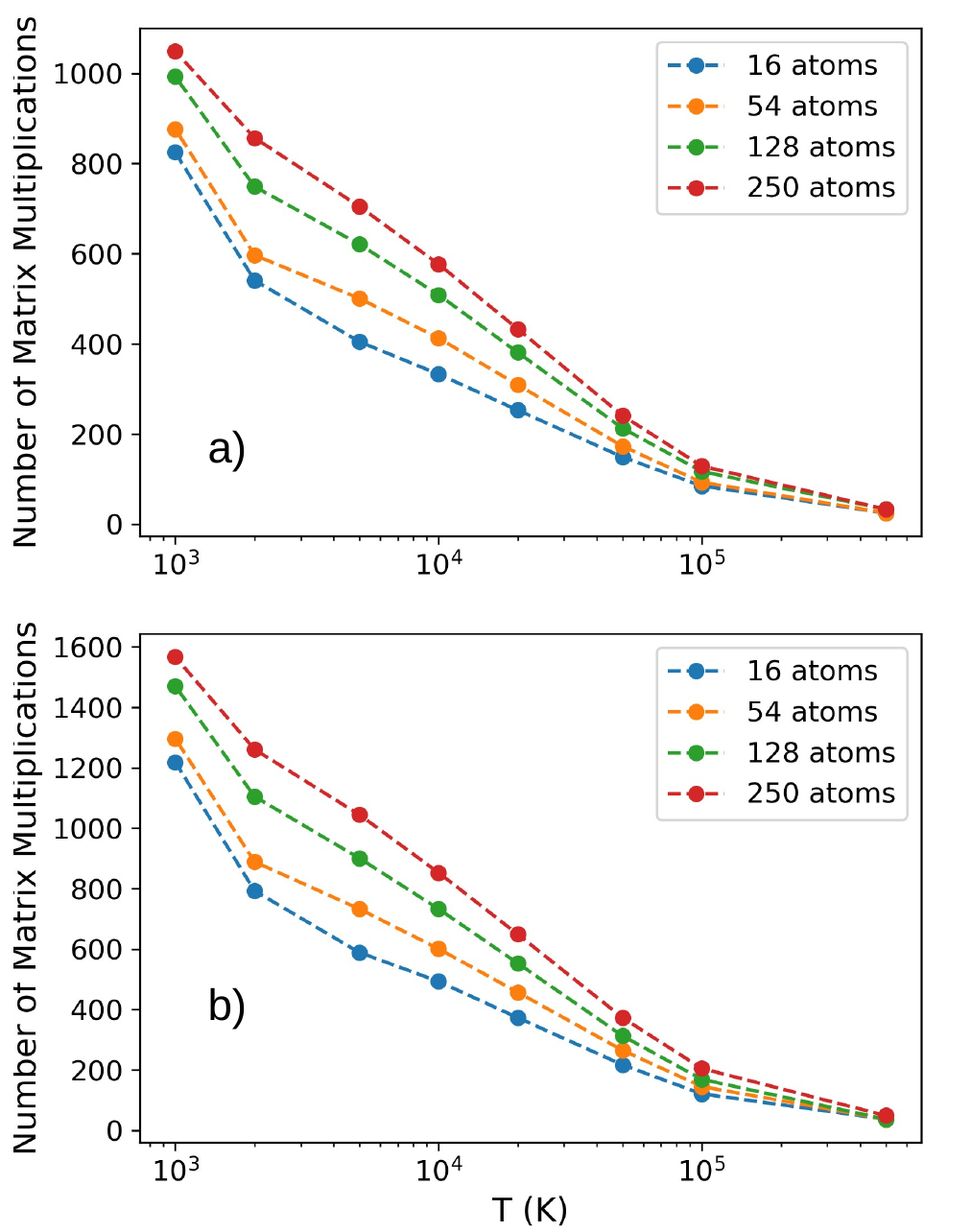}
    \caption{Number of matrix multiplications required to reach convergence for 16, 54, 128, and 250 atom supercells of aluminum as the temperature is changed when solving a) Eqs.~\eqref{eq:OmegaGC_deriv} and \eqref{eq:OmegaGC_iv} and b) Eqs.~\eqref{eq:OmegaSubbedCDeriv} and \eqref{eq:OmegaCIV}.  In each case, we used the adaptive implementation of our methods outlined in Algorithm~\ref{adaptive_algorithm}.
}
    \label{fig:steps_vs_size}
\end{figure}

We also tested how the adaptive step implementations of our method scale with the size of the system.  In this work, we consider systems of aluminum with 16, 54, 128, and 250 atoms.  The number of evaluations of Eq.~\eqref{eq:OmegaGC_deriv} or Eq.~\eqref{eq:OmegaCDeriv} required for our methods to reach convergence at different temperatures for each system are displayed in Fig.~\ref{fig:steps_vs_size}.  Each evaluation of Eq.~\eqref{eq:OmegaGC_deriv} requires four matrix multiplications and each evaluation of Eq.~\eqref{eq:OmegaSubbedCDeriv} requires six matrix multiplications.  If $S = 1$, such as when using an initial Cholesky or L{\"o}wdin orthogonalization of H, this reduces to three matrix multiplications per evaluation for either equation.  These results demonstrate that the number of calculations needed by the adaptive step implementations of our methods grows slowly as the system is increased from 16 to 250 atoms; in Supplementary Information II, we plot the number of multiplications vs. number of atoms (up to 1024) to show how the number of multiplications plateaus.  Furthermore, the linear relation between the number of matrix multiplications and a logarithmic scale of the temperature in Fig.~\ref{fig:steps_vs_size} indicates that the number of calculations required for convergence decreases exponentially with increasing temperature.

\begin{figure}
    \centering
    \includegraphics[width=\linewidth]{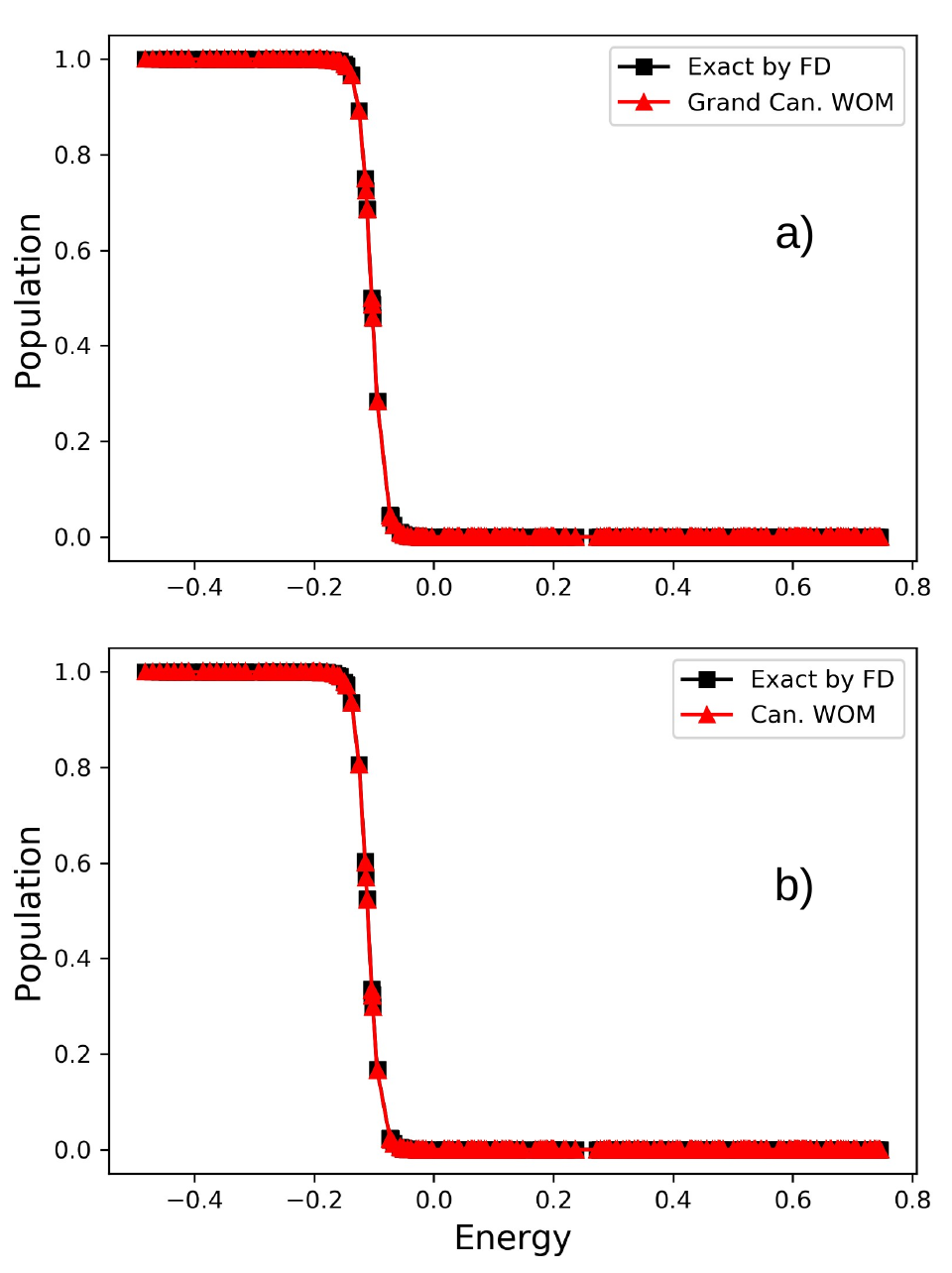}
    \caption{Eigenvalue spectra of a) the exact matrix obtained by calculating Eq.~\eqref{non_orth_fd} and the matrix obtained by solving Eqs.~\eqref{eq:OmegaGC_deriv} and \eqref{eq:OmegaGC_iv}; b) the exact matrix obtained by calculating Eq.~\eqref{eq:c_non_orth_fd} and the matrix obtained by solving Eqs.~\eqref{eq:OmegaSubbedCDeriv} and \eqref{eq:OmegaCIV} for a tight-binding treatment of aluminum with 250 atoms and 750 electrons.  In each case, the spectra of our results demonstrate strong agreement with the exact spectra.}
    \label{fig:al_eigs}
\end{figure}

\begin{figure}
    \centering
    \includegraphics[width=\linewidth]{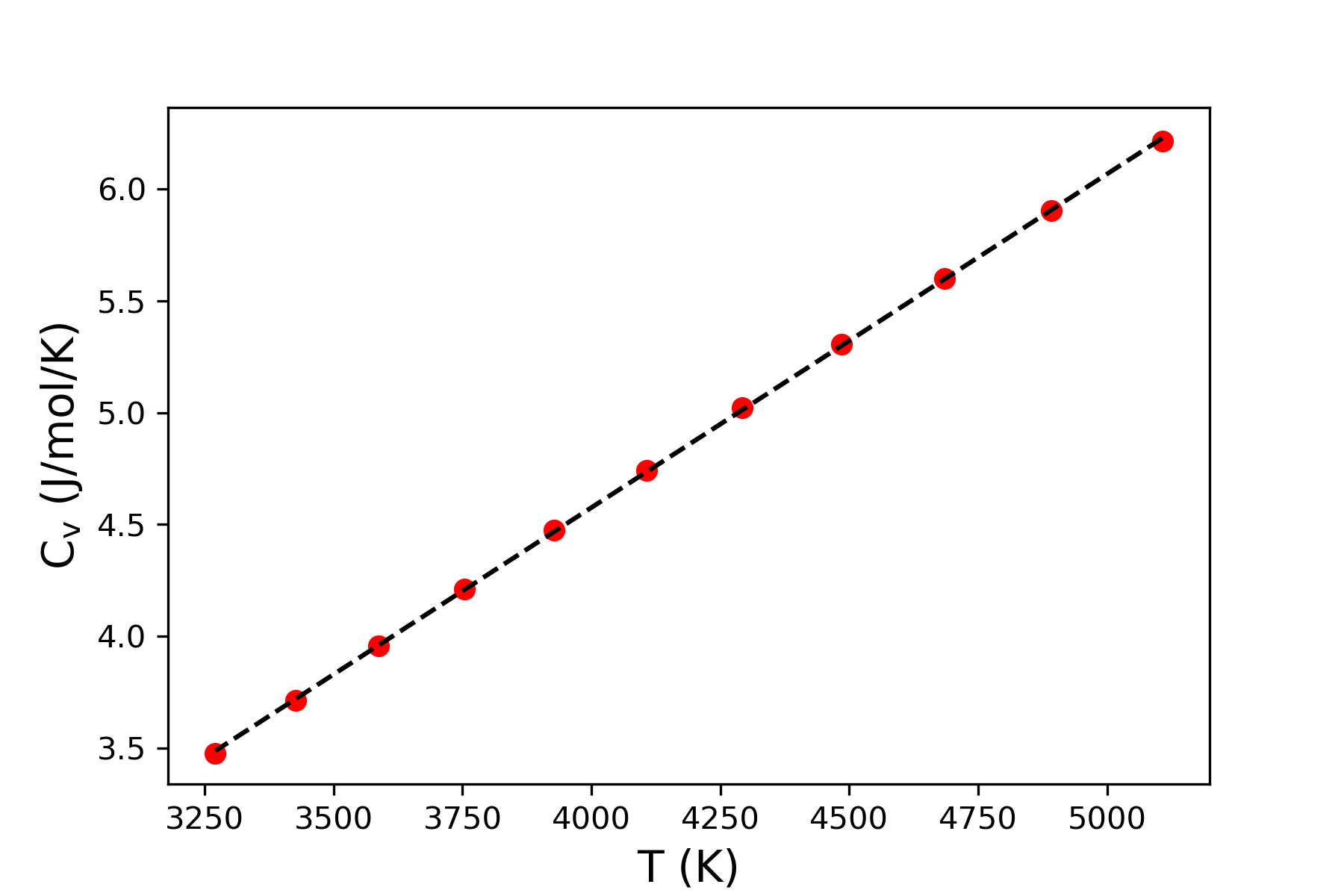}
    \caption{The dependence of the specific heat $C_v$ on the temperature $T$ for an aluminum system with 250 atoms.  The red dots are calculated according to Eq.~\eqref{eq:cv} and the black dashed line indicates the fit using a least squares linear regression.} 
    \label{fig:c_v_t}
\end{figure}

We also wanted to confirm that our canonical method can be used to obtain the correct trend of the specific heat $C_v$ due to the electrons while solving Eqs.~\eqref{eq:OmegaSubbedCDeriv} and \eqref{eq:OmegaCIV}, which can be obtained from
\begin{align}
    C_v = -2k\beta^2\Tr\left(S^{-1}\frac{d\Omega^\dag}{d \beta}\Omega A + S^{-1}\Omega^\dag \frac{d\Omega}{d\beta}A\right). \label{eq:cv}
\end{align}
At temperatures much lower than the Fermi temperature, $C_v \sim \gamma T$ where $\gamma$ is a constant that depends on the material.  For metals, the theoretical value $\gamma_{th}$ can be calculated using the Sommerfeld free electron theory, though there is often a discrepancy between $\gamma_{th}$ and the value observed in experiments $\gamma_e$.  The values for aluminum are $\gamma_{th} = 0.912 ~\mathrm{mJ/mol/K}^2$ and $\gamma_e = 1.35 ~\mathrm{mJ/mol/K}^2$~\cite{kittel_solid}.  To calculate the $\gamma$ obtained using our canonical method, we evaluated Eq.~\eqref{eq:cv} at several temperatures and performed a least squares linear regression to obtain the slope.  The data and resulting fit for the 250 atom supercell are plotted in Fig.~\ref{fig:c_v_t}.  We see that the fit had $R^2 = 0.99993$ and gave $\gamma = 1.493 \mathrm{mJ/mol/K^2}$, which is 10.6\% larger than $\gamma_e$.

\begin{figure}
   \centering
   \includegraphics[width=\linewidth]{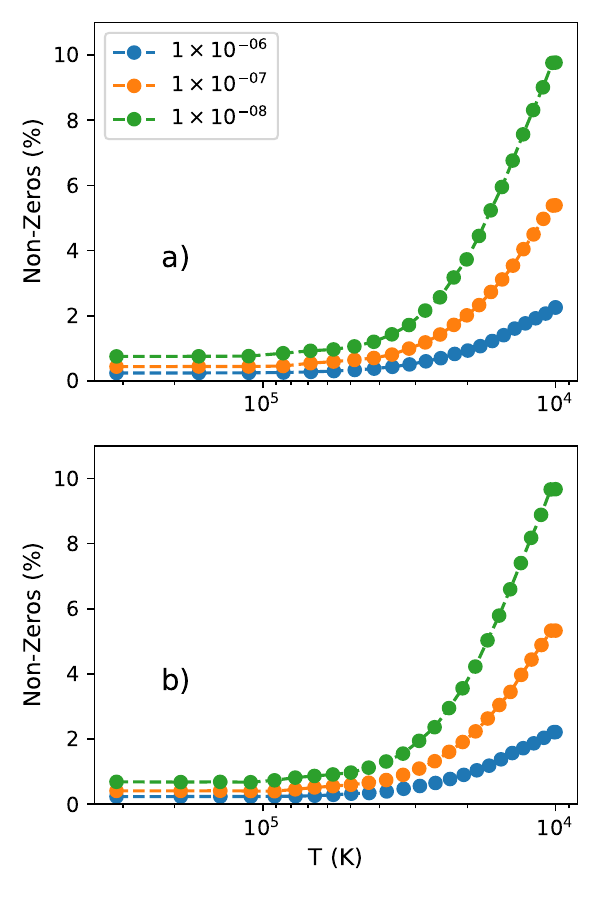}
   \caption{Sparsity of the density matrix of a 6912 atom supercell of silicon vs temperature. Calculations are performed for the: a) grand canonical and b) canonical ensembles. The legends indicate the threshold used when counting the number of elements in the density matrix that were non-zero.}
   \label{fig:sparsity}
\end{figure}

As discussed at the beginning of Sec.~\ref{sec:fin_temp}, sparsity arises naturally in electronic structure problems because of Kohn's near-sightedness principle and minimization based approaches to calculating the density matrix take advantage of this to achieve $O(N)$ scaling.  Because of this, we wanted to investigate the effect that temperature had on the sparsity of the density matrix. To do so, we implemented the WOM method in the NTPoly library \cite{dawson_massively_2018}. We evolve the L{\"o}wdin orthogonalized version of $\Omega$ in the grand canonical and canonical ensembles. As a test case, we used the tight-binding Hamiltonian of a bulk silicon system of 6912 atoms generated with the pbc-0-3 parameter set~\cite{sieck2000structure}. During the calculation, NTPoly automatically filters matrix values below some threshold to maintain sparsity. We plot the percentage of non-zero elements vs temperature for threshold choices of $10^{-6}$, $10^{-7}$, and $10^{-8}$ in Fig.~\ref{fig:sparsity}. With the WOM method, fill in of the density matrix occurs gradually as the temperature decreases, ensuring that sparse linear algebra can be used for the whole process. This, combined with the number of multiplications being roughly constant with the system size (Supplementary Information II), means that linear scaling performance can be achieved when the WOM method is used with sufficiently large systems.

\subsection{Comparison with the Fermi Operator Expansion}

\begin{figure}
   \centering
   \includegraphics[width=\linewidth]{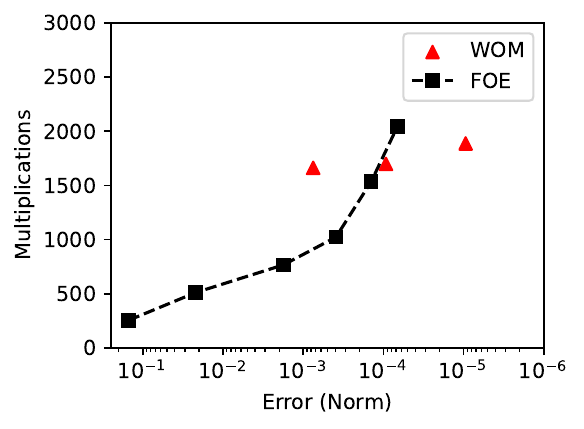}
   \caption{Number of multiplications require for the WOM and FOE methods to compute a 108 atom silicon system at 3157K in the grand canonical ensemble. WOM calculations were performed with a threshold of 0.1, 0.05, and 0.01. Error is reported as norm of the difference between the density matrix computed with each method and with exact diagonalization. In both cases, we use the L{\"o}wdin orthogonalized version of the Hamiltonian.}
   \label{fig:foe_comp}
\end{figure}

To further evaluate the WOM method, we compare it to the Fermi Operator Expansion (FOE) method, which expands the Fermi-Dirac distribution using Chebyshev polynomials~\cite{PhysRevB.51.9455}. The FOE method is implemented in the CheSS library~\cite{mohr_efficient_2017} and is regularly used in production calculations with the BigDFT code~\cite{Ratcliff2020}. As a test case, we use a 108 atom supercell of bulk silicon coming from the Siesta code~\cite{Garcia2020} with a final temperature of 3157K. We computed this system with the PBE functional~\cite{10.1103/PhysRevLett.77.3865}, the DZP basis set, a mesh cutoff of 300 Ry, an energy shift of 0.01 Ry, and ONCV pseudopotentials~\cite{10.1103/PhysRevB.88.085117} taken from PseudoDojo~\cite{vanSetten2018, Garcia2018}. The number of matrix multiplications and error in the density matrix using different tolerances for the WOM method and numbers of polynomials for FOE are plotted in Fig.~\ref{fig:foe_comp}. For this test case, we find that WOM and FOE require a similar number of multiplications, depending on the required precision.

There are a number of weaknesses of the FOE method that are addressed by WOM. First, the number of polynomials required depends heavily on the ratio of the HOMO-LUMO gap to the spectral width~\cite{mohr_efficient_2017}. For this reason, we have chosen a matrix coming from Siesta instead of the earlier tight-binding matrices as a test case. The number of polynomials required would be even larger if, for example, core electrons were included. In contrast to FOE, the ratio of the gap to spectral width only affects the starting temperature (after the first step) of the WOM iterations. In Supplementary Information III, we investigate more closely how the number of iterations is affected by the size of the gap and spectral width. Another drawback to the FOE method is that to compute the chemical potential, one needs to store all intermediate matrix powers in memory, whereas WOM only needs to store a few matrices. Storing intermediate matrices would also be necessary if one wanted to compute the density matrix at different temperatures for properties like the specific heat, which come out automatically with WOM. This characteristic also makes WOM potentially useful when restarting a calculation: the stored wave operator matrix from a previous calculation can be loaded in from disk and iterated to an even lower temperature.

In order to reduce the number of multiplications required for FOE, a variation of the Patterson - Stockmeyer method~\cite{Paterson1973} was proposed by Liang and coworkers~\cite{Liang2003}. However, this method is challenging to use in practice: there is significant fill in that occurs with intermediate matrices and root finding is needed to get the chemical potential (though some intermediate matrices can be saved). This method has recently been shown to be useful in the dense matrix case using GPUs, where managing sparsity is not an issue~\cite{Finkelstein2023}. We note as well that evaluations of the Fermi-Operator Expansion frequently target the zero temperature case, where the number of polynomials can be reduced by expanding the complementary error functions instead of the true Fermi-Dirac distribution or by using the Jackson approximation~\cite{Jay1999}. For the zero temperature case, these modifications may be used to help the FOE method significantly outperform WOM (see Supplementary Information III for more discussion of the zero temperature case).

\section{Conclusion}\label{sec:conclusion}

We have developed positivity-preserving methods for calculating the Fermi-Dirac density matrices at finite temperatures for both the grand canonical (Sec.~\ref{sec:gc_fin_temp}) and canonical (Sec.~\ref{sec:c_fin_temp}) ensembles. These methods are based on minimization of the square root of the density matrix ($\Omega$) and the modelling of a physical process of cooling a state initially at infinite temperature down to the temperature of interest.  The effectiveness of WOM is demonstrated by considering tight-binding models of aluminum and silicon of various system sizes.  Additionally, the number of steps required to reach convergence is shown to be independent of the number of atoms in the system.  To the best of our knowledge, this work is the first to make use of $\Omega$ as a computational tool.  This approach may lay the groundwork for a new generation of methods that utilize $\Omega$ for solving electronic structure problems. To facilitate this, we made the code publicly available on GitHub \cite{leamer_git}.  For a practical implementation in Density Functional Theory codes, using minimization as a replacement for the self-consistent cycle could be used to further reduce the computational cost~\cite{Salek2007,Flamant2019}. A possible future direction is to utilize the fact that a low-temperature Fermi-Dirac density matrix is of a low rank, hence it is possible to further accelerate calculations by utilizing low-rank corner space techniques recently developed for solving master equations for large open quantum systems~\cite{le_bris_low-rank_2013, finazzi_corner-space_2015, mccaul_fast_2021, chen_low_2020, donatella_continuous-time_2021}.

\begin{acknowledgments}

The authors are grateful to George Booth for helping with the PySCF library \cite{PYSCF} and valuable discussions. J.M.L. was supported by the Louisiana Board of Regents’ Graduate Fellowship Program. D.I.B. is supported by the Army Research Office (ARO) (grant W911NF-23-1-0288; program manager Dr.~James Joseph), Air Force Office of Scientific Research (AFOSR) Young Investigator Research Program (grant FA9550-16-1-0254; program manager Dr.~Fariba Fahroo), and the Alexander von Humboldt Foundation (Humboldt Research Fellowship for Experienced Researchers).

\end{acknowledgments}

\section*{Conflict of Interest}

The authors have no conflicts to disclose.

\section*{Data Availability Statement}

Implementations of the described algorithms are publicly available on GitHub \cite{leamer_git}. An implementation of the described algorithms is also available as part of NTPoly~\cite{ntpoly_git}. The remaining data that support the findings of this study are available from the corresponding author upon reasonable request.

\appendix

\section{Non-Orthonormal Fermi-Dirac}\label{app:NonOrthonormalFD}
We want to derive an expression for the Fermi-Dirac distribution that is valid for a general, non-orthonormal basis.  Consider the general, non-orthonormal basis set: $\{\ket{i}\}_{i=0}^N$ where $N$ is the dimension of the system in question.  Let $S = (\left< i \middle| j \right>)$ be the overlap matrix, $\hat{H}$ be the Hamiltonian operator, and $\hat{\rho}$ be the  Fermi-Dirac distribution
\begin{equation}
    \hat\rho(\beta) = \frac{\hat{I}}{\hat{I} + \exp[\beta(\hat{H}-\mu \hat{I})]}.
\end{equation}
Now let $P = (\braket{i|\hat\rho|j})$ and $H = (\braket{i|\hat{H}|j})$ be the representations of $\hat\rho$ and $\hat{H}$ in the non-orthonormal basis, respectively.  The Taylor series expansion for $\hat\rho$ reads
\begin{equation}
    P = \sum_{n=0}^\infty a_n \braket{i|\hat{A}^n|j},
    \label{P}
\end{equation}
where $\hat{A} = (\hat{H}-\mu)$ and has the representation $A = (\braket{i|\hat{A}|j})$ in the non-orthonormal basis.  The goal is to find the representation of $\braket{i|\hat{A}^n|j}$ in the non-orthonormal basis so that we can obtain a concise expression for $P$. We start by considering how $\hat{A}$ acts on a basis vector $\ket{j}$.  If we expand $\ket{j}$ in the basis $\{\ket{k}\}$, we get
\begin{align}
    \braket{i|\hat{A}|j} = \braket{i|\sum_k C_{kj}|k} 
    = \sum_k C_{kj}\braket{i|k} 
            = SC
         \label{a_on_j}
\end{align}
where $C = S^{-1}A$ is just the matrix of coefficients obtained from $\hat{A}\ket{j}$.  We can perform similar expansions to obtain $\braket{i|\hat{A}^{n}|j}$ for $n > 1$ and then use induction to show that
\begin{equation}
    \braket{i|\hat{A}^n|j} = A(S^{-1}A)^{n-1}.
\end{equation}
If we multiply both sides of Eq.~\eqref{P} by $S^{-1}$ on the left and substitute in for $\braket{i|\hat{A}^n|j}$, we obtain
\begin{align}
    S^{-1}P &= \sum_{n=0}^\infty a_n(S^{-1}A)^{n}
        = \left[I + \exp[\beta S^{-1}A]\right]^{-1}.
\end{align}
Finally, we replace $A$ with $\braket{i|(\hat{H}-\mu I)|j}$ and solve for $P$ to obtain
\begin{equation}
    P = S\left[I + \exp[\beta(S^{-1}H-\mu I)]\right]^{-1}.
\end{equation}
Note that $P$ is the real density matrix in the non-orthogonal basis.  Electronic structure codes frequently make use of the density kernel $K$, which is the representation of the density matrix in the dual basis \cite{mohr_daubechies_2014, haynes_density_2008}.  The density kernel can be obtained by
\begin{align}
    K = S^{-1}PS^{-1},
\end{align}
which is the transformation of $P$ to the dual basis.

\section{The root of P}\label{app:omega_proof}

We want to derive Eq.~\eqref{eq:sqrt_P}.  To simplify matters, we introduce the following
\begin{align}
    f(x) &= [1+e^{x}]^{-1/2}, \\
    A &= S^{-1}H-\mu I, \label{eq:Adef}
\end{align}
so that Eq.~\eqref{eq:sqrt_P} can be written as
\begin{align}
    P(\beta) &= Sf(\beta A)f(\beta A), \\
    \Omega(\beta) &= S^{1/2}f(\beta A).
\end{align}
Now consider expressing $\Omega^\dag\Omega = \Omega(\beta)^\dag\Omega(\beta)$ in terms of the Maclaurin series expansion of $f(\beta A)$, which yields
\begin{align}
    \Omega^\dag\Omega =& f(\beta A^\dag)Sf(\beta A), \\
    =& \sum_{i,j=0}^\infty \frac{f^{(i)}(0)}{i!}(A^\dag)^i\beta^i S \frac{f^{(j)}(0)}{j!}A^j\beta^j.
\end{align}
Grouping by powers of $\beta$ gives
\begin{align}
    \Omega^\dag\Omega &= \sum_{k=0}^\infty \sum_{m=0}^k\frac{f^{(m)}(0)f^{(k-m)}(0)}{m!(k-m)!}(A^\dag)^l S A^{k-m}\beta^k.
\end{align}
It follows from Eq.~\eqref{eq:Adef} that $SA = (SA)^\dag = A^\dag S$, which can be used with the fact that $\binom{k}{m} = \frac{k!}{(k-m)!m!}$ to obtain
\begin{align}
    \Omega^\dag\Omega = \sum_{k=0}^\infty \frac{1}{k!}\sum_{m=0}^k \binom{k}{m}f^{(m)}(0)f^{(k-m)}(0)SA^k\beta^k. \label{eq:prod_rule}
\end{align}
Note now that
\begin{align}
    (f(\beta A)f(\beta A))^{(k)}|_{\beta=0} = \sum_{m=0}^k\binom{k}{m}f^{(m)}(0)f^{(k-m)}(0)A^k,
\end{align}
which can be substituted into Eq.~\eqref{eq:prod_rule} to obtain
\begin{align}
    \Omega^\dag\Omega &= \sum_{k=0}^\infty \frac{1}{k!}S\left[f(\beta A)f(\beta A)\right]^{(k)}\big|_{\beta=0}\beta^k, \\
      &= \sum_{k=0}^\infty \frac{1}{k!} P^{(k)}(0)\beta^k, \\
      &= P(\beta).
\end{align}

\hfill \break

\bibliography{main}

\end{document}


\preprint{AIP/123-QED}

\title{Supplementary Information - Positivity Preserving Density Matrix Minimization at Finite Temperatures via Square Root}
\author{Jacob Leamer}
\affiliation{Department of Physics and Engineering Physics, Tulane University, 6823 St. Charles Ave., New Orleans, LA 70118, USA}
\author{William Dawson}
\affiliation{ 
RIKEN Center for Computational Science, Kobe, Hyogo,
650-0047, Japan.
}%
\author{Denys I. Bondar}
\affiliation{Department of Physics and Engineering Physics, Tulane University, 6823 St. Charles Ave., New Orleans, LA 70118, USA}

\maketitle

\section{Error in Occupation}

\begin{figure}
    \centering
    \includegraphics[width=\linewidth]{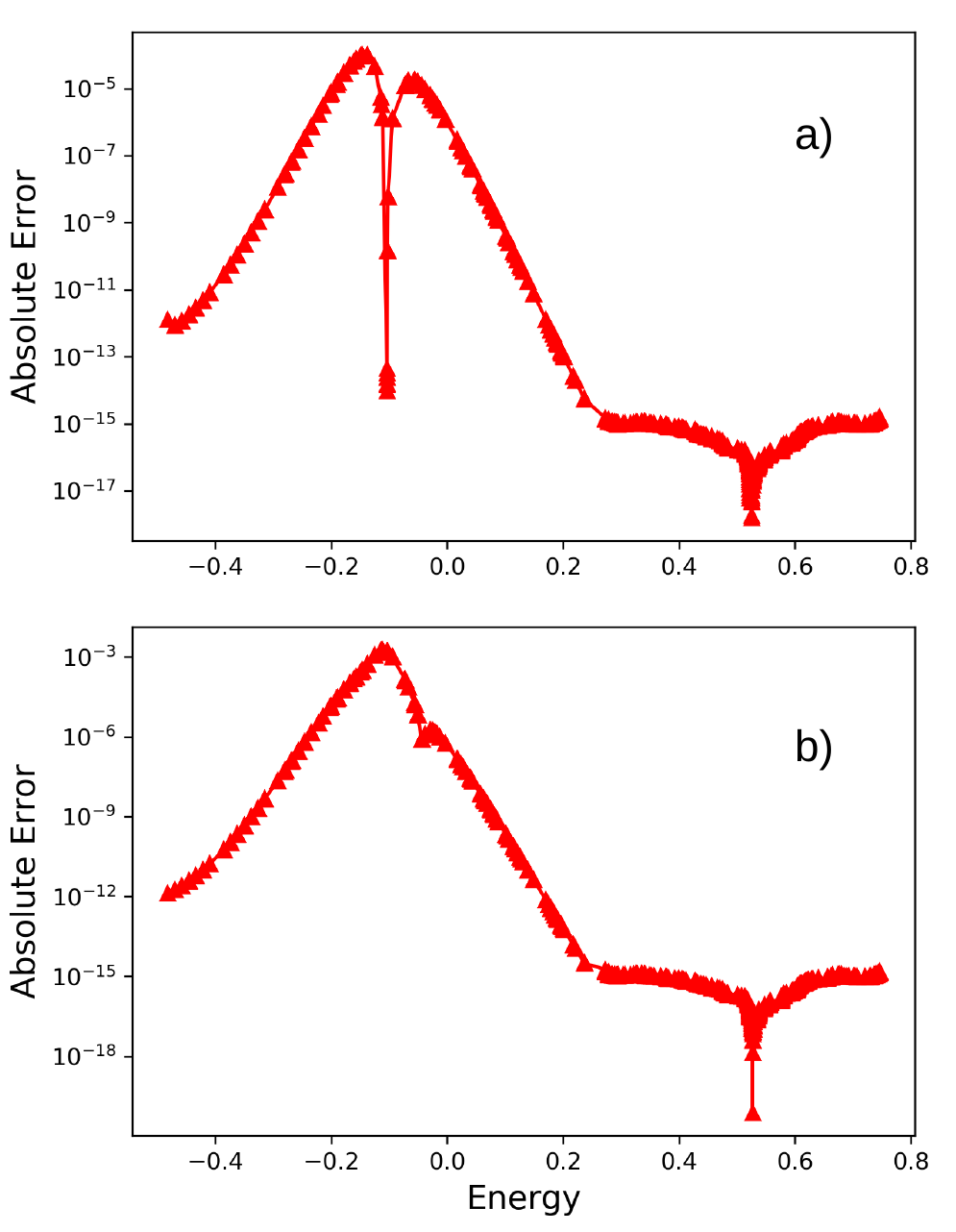}
    \caption{Absolute error in the occupation numbers between the results obtained by WOM and exact diagonalization for the a) grand canonical and b) canonical ensembles.}
    \label{fig:error_vs_energy}
\end{figure}

The error in the occupation numbers of the resulting density matrices obtained by WOM vs. diagonalization for the 250 atom aluminum tight binding system is plotted in Fig.~\ref{fig:error_vs_energy}. The error is largest near the Fermi level; however, the errors are sufficiently low for practical calculations.

\section{Multiplications vs. System Size}

\begin{figure}
    \centering
    \includegraphics[width=\linewidth]{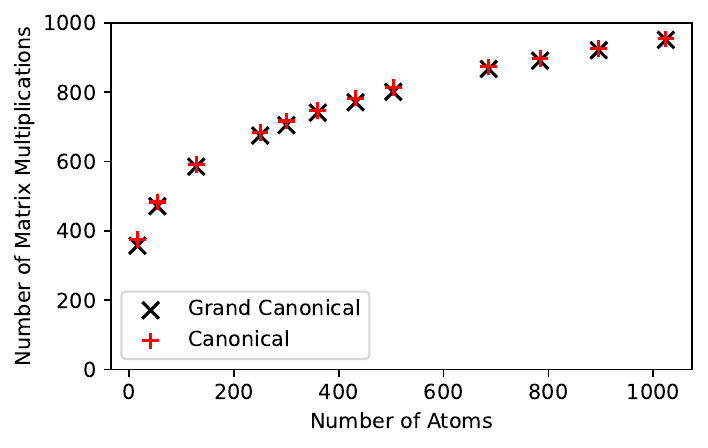}
    \caption{Number of multiplications required to reach 3157K in the grand canonical and canonicals.}
    \label{fig:mvsa}
\end{figure}

We examined how the number of multiplications grows with the number of atoms using the aluminum tight binding systems as an example (Fig.~\ref{fig:mvsa}). For this analysis, we use the L{\"o}wdin orthogonalized Hamiltonian for computational efficiency. We find that the number of multiplications initially grows with the system size, before finally saturating.

\section{Impact of the HOMO-LUMO Gap}

\begin{figure}
    \centering
    \includegraphics[width=0.8\linewidth]{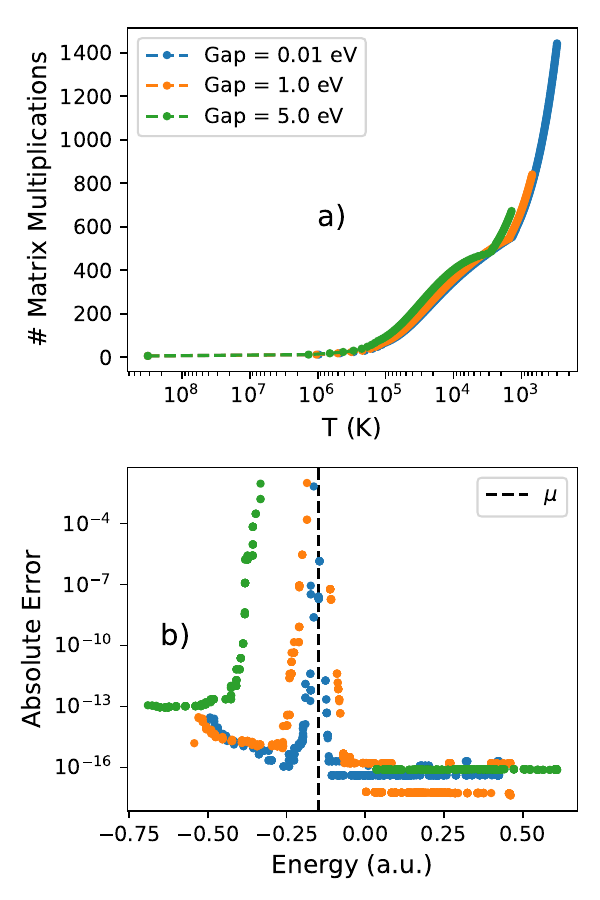}
    \caption{The effect of the size of the gap on the WOM method. a) The number of matrix multiplications vs. temperature. b) Absolute error in the occupancy vs. orbital energy. Calculations are performed with the L{\"o}wdin orthogonalized Hamiltonian with a target temperature of 300K.}
    \label{fig:artificial_gap}
\end{figure}

In order to understand the impact of the size of the HOMO-LUMO gap on the convergence of the WOM method, we used the tight-binding model of silicon (108 atoms) and artificially modified the gap by shifting the occupied and virtual orbitals towards / away from the chemical potential. In Fig.~\ref{fig:artificial_gap}, we plot the number of multiplications vs. temperature and the errors in occupancy achieved with the Grand Canonical Ensemble version of WOM using several different values for the HOMO-LUMO gap. We find that the number of iterations and error is fairly insensitive to the gap. One significant difference comes from changes in the spectral width (from shifting the eigenvalues outward), which changes the initial temperature after the first step. Additionally, when a low temperature solution is sought, iterations can exit early for systems with a larger gap. This is because the difference in the wave operator at a lower temperature is close to the ground state, which triggers the early exit condition of Algorithm 1. From this data, we conclude that WOM can be applied to both insulating and metallic systems without any changes.

\bibliography{supplementary}